# YAG nano-light sources with high Ce concentration

**B. Masenelli[1], O. Mollet[2], O. Boisron[3], B. Canut[1], G. Ledoux[3], J.-M. Bluet[1], P. Mélinon[3], Ch. Dujardin[3], S. Huant[2]**

[1] Institut des Nanotechnologies de Lyon, UMR 5270 CNRS and INSA Lyon, 7 avenue Jean Capelle, Université de Lyon 69621 Villeurbanne cedex, France

[2] Institut Néel, CNRS and Université Joseph Fourier, BP 166, 38042 Grenoble, France

[3] Institut Lumière Matière, UMR5306 Université Lyon 1-CNRS, Université de Lyon 69622 Villeurbanne cedex, France

E-mail : bruno.masenelli@insa-lyon.fr

**Abstract.** We investigate the luminescence properties of 10 nm YAG nanoparticles doped with Ce ions at 0.2%, 4% and 13% that are designed as active probes for Scanning Near field Optical Microscopy. They are produced by a physical method without any subsequent treatment, which is imposed by the desired application. The structural analysis reveals the amorphous nature of the particles, which we relate to some compositional defect as indicated by the elemental analysis. The optimum emission is obtained with a doping level of 4%. The emission of the YAG nanoparticles doped at 0.2% is strongly perturbed by the crystalline disorder whereas the 13% doped particles hardly exhibit any luminescence. In the latter case, the presence of $Ce^{4+}$ ions is confirmed, indicating that the Ce concentration is too high to be incorporated efficiently in YAG nanoparticles in the trivalent state. By a unique procedure combining cathodoluminescence and Rutherford backscattering spectrometry, we demonstrate that the enhancement of the particles luminescence yield is not proportional to the doping concentration, the emission enhancement being larger than the Ce concentration increase. Time-resolved photoluminescence reveals the presence of quenching centres likely related to the crystalline disorder as well as the presence of two distinct Ce ions populations. Eventually, nano-cathodoluminescence indicates that the emission and therefore the distribution of the doping Ce ions and of the defects are homogeneous.

## 1. Introduction

Yttrium aluminium garnet (YAG), of formula $Y_3Al_5O_{12}$, is a material that can be doped with rare earth ions in substitution for Y and commonly used in many applications (lasers, refractive coatings, mercury vapour lamps, cathode ray tubes, scintillators…) [1,2,3]. Among the doped YAG materials, Ce doped YAG nanoparticles are subject of intensive research motivated essentially by their use in white light emitting diodes (LEDs) [4,5,6,7]. In such an application, the blue emission of a high brightness GaN or InGaN LED is partially absorbed by the $Ce^{3+}$ ions of the YAG nanoparticles and converted by photoluminescence process into yellow light (peaked at 550 nm approximately). More recently, such nano-objects have been proposed as efficient nano light-emitters [8] for photonics with good properties with respect to bleaching and blinking.

Whatever the application foreseen, it has been sought to improve the external luminescence efficiency of Ce:YAG nanoparticles through the increase of the $Ce^{3+}$ concentration. When dealing with single nanoparticles, absorption of the incoming laser beam becomes a major issue. Therefore, increasing the doping content, even at higher level than it is regularly admitted for optimized phosphors has to be considered. In bulk YAG, a high concentration of Ce (typically over 2% atomic concentration with respect to Y) is not necessarily beneficial for it leads first to a luminescence quenching due to interaction between neighbour $Ce^{3+}$ ions. Second, with too high a Ce concentration the cubic structure of YAG is no longer stable with respect to the perovskite structure and this change is harmful to the luminescence. However, several recent studies have investigated the effect of a high concentration of Cerium in YAG nanoparticles. It was postulated that turning to nanoparticles could help to accommodate the excess $Ce^{3+}$ ions thanks to the proclivity of nanoparticles to relax the stress induced by the high doping concentration. Haranath and co-workers [6] synthesized Ce:YAG nanoparticles of approximately 40 nm in diameter by single-step auto-combustion process with Ce composition from 0.3 to 30% in substitution for Y. According to their work, the incorporation of Ce in YAG nanoparticles is possible up to 10%. The optimum concentration as far as luminescence is concerned was 6 % and the incorporation was accompanied with a slight red shift of the Ce emission spectrum. On the other hand, Okuyama *et al.* [9] have found an optimum concentration of 4% in nanoparticles with identical size while Poitevin *et al.* [10] have obtained the best luminescence yield with a concentration of 0.5%. Still more controversial, Tok and co-workers [11] observed a blue shift of the Ce emission in 30 nm YAG nanoparticles. As expected, because of the increased presence of defects in nanoparticles, the luminescence efficiency of the nanoparticles is always lower than the bulk one. Haranath and co-workers [6] suggested that further reducing the size of the particles to the quantum confinement regime could improve the quantum efficiency. For this purpose, synthesizing particles of diameter of 10 nm or less is a major issue. Isobe and co-workers [12] managed to synthesize 10nm crystallized Ce:YAG nanoparticles doped at 1%. However, to the best of our knowledge, studies dealing with such small nanoparticles are scare and the effect of the doping concentration on their luminescence efficiency has not been investigated.

The aim of the present study is thus to synthesize Ce:YAG nanoparticles of 10 nm diameter and probe the efficiency of high doping levels. Indeed, when they are used in nano-optics in the single particle configuration, reaching high doping level of active ions is a key aspect in order to improve the absorption coefficient, which is rather low as compared to semiconductor quantum dots. Nevertheless, in doped oxides, increasing the doping concentration generally leads to non-radiative processes inducing a decrease of the luminescence yield. However, contrary to other studies targeting applications in white light LEDs, we seek to avoid high temperature annealing which is known to improve the crystallinity but may be harmful for further use of the nanoparticles as active probes in SNOM (scanning near field optical microscopy) [8,13,14]. For this purpose, nanoparticles doped with cerium at 0.2, 4 and 13 % have been produced by a physical way, the LECBD (low energy cluster beam deposition) technique, and their structural and optical properties have been investigated. We have first observed that contrary to other rare earth oxide nanoparticles produced by the same method, the YAG ones are amorphous which consequently affects their emission properties. Nevertheless, the incorporation of 4% Ce is possible and leads to an optimal luminescence yield under cathodo-excitation. Moreover, the increase in the luminescence yield is not proportional to the doping level and related to the presence of quenching centres. The presence of quenching is further confirmed by time-resolved photoluminescence spectroscopy.

**2. Experimental**
The nanoparticles have been synthesized by the aforementioned LECBD technique that has been described in details elsewhere [15]. Briefly, a target pellet is made by the sintering of a mixture of the different metal oxides ($CeO_2$, $Y_2O_3$, $Al_2O_3$) in the desired proportions. The pellet is ablated by a pulsed Nd:YAG laser (532 nm, 10Hz repetition rate) creating a plasma. The plasma is cooled by a continuous flow of a buffer gas (He, 25 mbar) triggering the formation of nucleation embryos (dimmers, trimmers). The formation of the nanoparticles is obtained by accretion of the plasma species during a subsequent adiabatic expansion (~$10^8$ K/s cooling rate). The expansion occurs when the mixture flows through a micrometer nozzle separating the nucleation chamber at high pressure (25 mbar) from the deposition chamber at low pressure ($10^{-7}$ mbar). The resulting nanoparticles are subsequently deposited as thin films (about one hundred nanometers thick) on silicon substrates without destruction, since the kinetic energy per atom is lower (by one order of magnitude) than the binding energy per atom among the particle (cf. Figure 1).

The elemental analysis of the nanoparticle films have been performed by X-ray Photoelectron Spectroscopy (XPS) and Rutherford Backscattering (RBS) spectrometry. The XPS data have been acquired using the Al $K_\alpha$ emission line at 1486.6 eV. The RBS analysis was performed using $^4He^+$ ions of 2 MeV energy delivered by the 4 MV Van de Graaff accelerator of the Nuclear Physics Institute of Lyon (IPNL). The backscattered particles were detected with a 13 keV resolution implanted junction set at an angle of 160° with respect to the beam axis.

The crystalline structure of the films has been investigated by X-Ray diffraction (XRD) analysis performed on a Bruckner D8 advance diffractometer with Cu $K_\alpha$ radiation in grazing incidence configuration ($\theta = 0.5°$). Transmission electron microscopy analysis either in the conventional mode (TEM) or in the high-resolution mode (HRTEM) has been performed on a TOPCON microscope at a 200 kV acceleration voltage to obtain the shape and size of the particles as well as their crystalline structure.

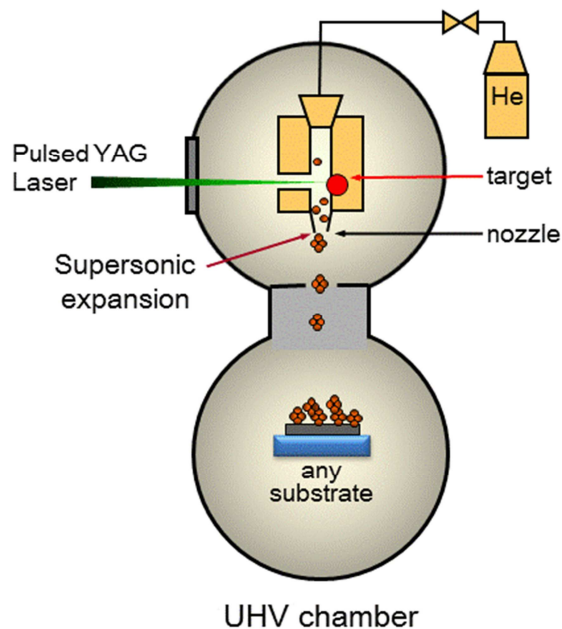

**Figure 1: scheme of the LECBD set-up. A plasma generated by the laser ablation of the target pellet undergoes a supersonic expansion through the nozzle, quenching the nucleation and leading to the formation of a beam of uncapped nanoparticles. The nanoparticles are subsequently deposited on any substrate in high or ultra high vacuum.**

The emission spectra of the nanoparticles have been acquired by a homemade cathodoluminescence (CL) set up. An electron gun (accelerating voltage up to 5keV) with a spot size of 1 mm$^2$ is used as the excitation source and a Jobin-Yvon Triax 320 spectrometer coupled to a Si CCD collects the light through an optical fibre. In order to quantitatively compare the emission of our films we have acquired their spectra in the same run, with exactly the same experimental conditions. This ensures that we avoided any measurement artefact, except possible change in light diffusion (which are unlikely since the roughness and texture of the films are identical) which may slightly affect the light collection. To be able to quantify the relative CL efficiency, we have measured by RBS the amount of matter that was present under each CL spot. The thickness of emitting matter was estimated by Monte Carlo simulations (casino code) to 120 nm at a 4.5 kV accelerating voltage. We have subsequently deduced the light yield emitted between 415 and 700 nm by a unit mass of Ce:YAG nanoparticles. The homogeneity of the light emission has also been probed at the nanoscale by a Gatan CL3 module on a field emission gun Tescan Mira 3 scanning electron microscope (SEM). The beam size was 7 nm and the acceleration voltage 30 keV resulting in a lateral resolution of approximately 50 nm. Eventually the luminescence decay of the particles has been probed using a doubled femtosecond Ti-Sapphire laser (Tsunami from Spectra-Physics) tuned at 450 nm. The out coming train pulse was filled to a pulse picker so as to obtain a frequency of 800 kHz. The emitted light was collected in an epifluorescence configuration and filtered at 580 nm ± 15 nm corresponding to the $Ce^{3+}$ emission. The light was detected by means of a cooled photomultiplier whose signal was fed to a Picoquant TimeHarp 200 time-correlated single photon counting system. The time resolution of the system is slightly better than 1ns.

**3. Results and discussions**

Figure 2 shows a typical TEM image of the deposited nanoparticles. It can be seen that the particles have no definite shape, some being compact, and others being more elongated. The elongated structures are likely to result from the sticking of compact particles as noticed in a previous study [8]. Considering the diameter of the compact particles and the width of the elongated ones, an average size of approximately 10 nm can be deduced. HRTEM imaging has been performed (cf. inset of Figure 2) but no crystalline particle has been observed, going to show that the particles are amorphous. The amorphous state of the nanoparticles assembled films is confirmed by the XRD analysis where no diffraction peak has been observed. This is in contrast with previous theoretical [16] and experimental studies [17], which proves that rare earth oxide nanoparticles of more simple formulation grown by the same method are crystalline provided that they are stoichiometric.

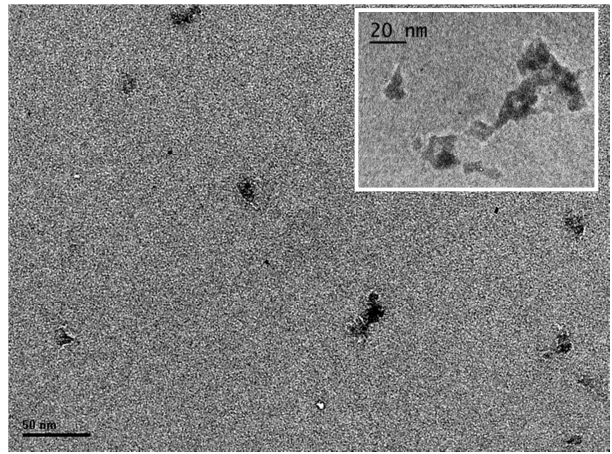

**Figure 2 : TEM image of Ce:YAG nanoparticles. The scale bar represents 50 nm. Inset: HRTEM image of some selected nanoparticles. The particles do not appear as crystallized**

The key issue regarding crystallinity in such ionic compounds is thus the chemical composition. The latter has been determined both by XPS and RBS. Table 1 presents the composition of the three batches of samples. Most of the samples have an excess of oxygen. This is probably due to OH or water adsorption on the particles surface exposed to air. It is indeed well known that oxides are highly sensitive to air moisture [18,19]. Nevertheless, the fact that the particles are not stoichiometric very likely favours the creation of crystalline defects and consequently the achievement of the amorphous state. In many studies, the synthesized samples are amorphous and do not exhibit any luminescence [9,10,20,21] if not annealed at several hundreds of degrees for several hours. In our strategy we seek to avoid post-deposition annealing, which not only is incompatible with the subsequent use of the particles as photonic nano-sources, but also can modify their size.

**Table 1: chemical composition determined by XPS and RBS for the three kinds of samples. The XPS and RBS compositions are given with a precision of 10 and 2% respectively.**

|  | Sample 1 | Sample 2 | Sample 3 |
|---|---|---|---|
| Chemical formulation | $Y_3Al_5O_{14.7}$ (RBS) $Y_3Al_5O_{13.7}$ (XPS) | $Y_3Al_5O_{16}$ (RBS) | $Y_{1.8}Al_5O_{10}$ (RBS) $Y_{2.5}Al_5O_{14}$ (XPS) |
| Ce/(Y+Ce) ratio | 0.2% | 4% | 13 (RBS) / 16% (XPS) |

For the samples with Ce concentrations of 0.2% and 4% the ratio Y/Al is stoichiometric whereas the samples with 13% Ce concentration have a deficit of Y. The XPS spectra (cf. Figure 3) of the constituting elements are in agreement with spectra reported for bulk crystalline YAG [19] meaning that despite the amorphous state of the particles the elements are in the appropriate chemical environment and hence have the correct valence state. Only for the samples with the highest content of Ce have we been able to collect the contribution of the Ce ions. In this case, a small amount of $Ce^{4+}$ ions [22,23] is detected along with the $Ce^{3+}$ contribution (see Figure 3). The total incorporation of 13% of Ce was not achieved and this likely leads to the presence of $CeO_2$ inclusions among the films in which Ce ions can adopt the 4+ valence.

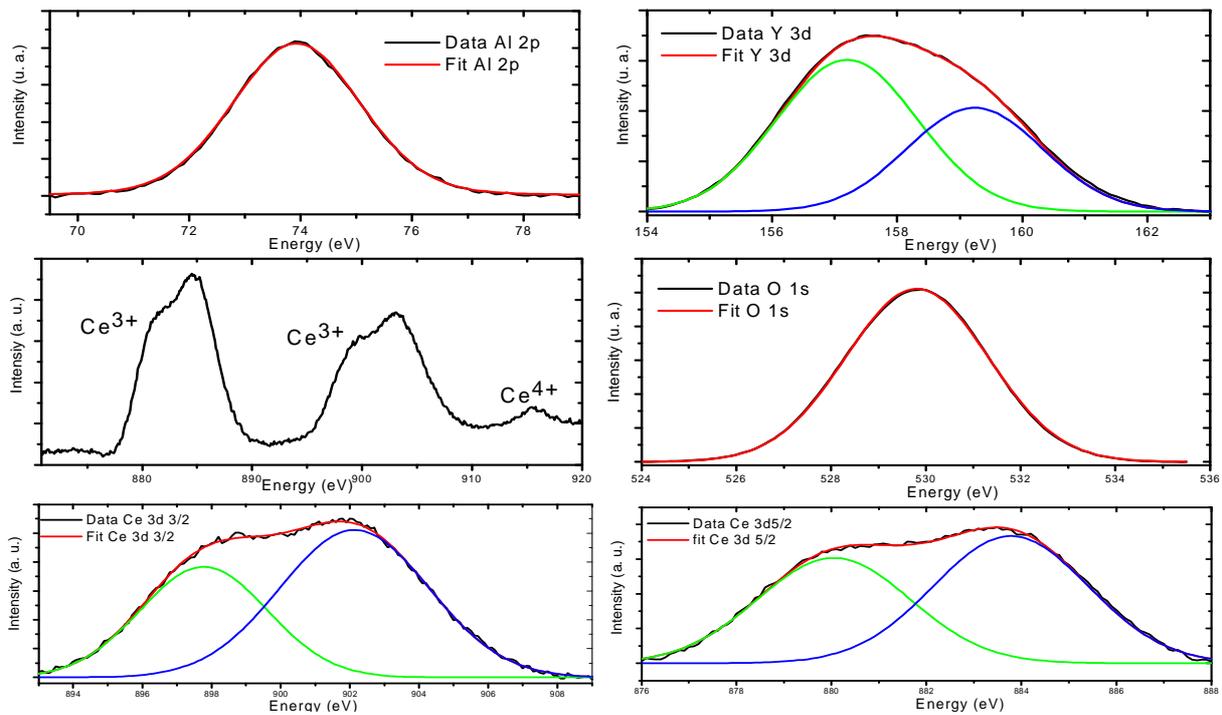

**Figure 3: XPS spectra of the elements constituting the 13% doped YAG nanoparticles along with the fitting curves. From top left to bottom right: Al2p, Y3d, survey of the Ce contributions, O1s, Ce3d 3/2 and Ce 3d 5/2 levels respectively. The Al 2p spectrum is actually made of two contributions, which are not resolved with our apparatus. The O1s peak is large as reported in the literature [19] revealing in particular the contribution of adsorbed species. The presence of $Ce^{4+}$ is confirmed on the survey spectrum along with the contributions of $Ce^{3+}$.**

Concomitantly, the samples doped at 13% with Ce have an inhomogeneous luminescence. Most parts of the samples do not show any light and only local regions emit a low emission at 400 nm characterized by an asymmetric shape (cf. Figure 4). This contribution can result from local variations of composition and its assignment is not clear at present. On the contrary, the emission of the YAG particles doped at 0.2% or 4% with Ce, shown on Figure 4, are representative of these batches of samples. The $Ce^{3+}$ emission from the 0.2% and 4% doped nanoparticles is identical to the emission of $Ce^{3+}$ ions in bulk YAG, in accordance with the previous XPS results revealing a local chemical environment close to the bulk YAG one.

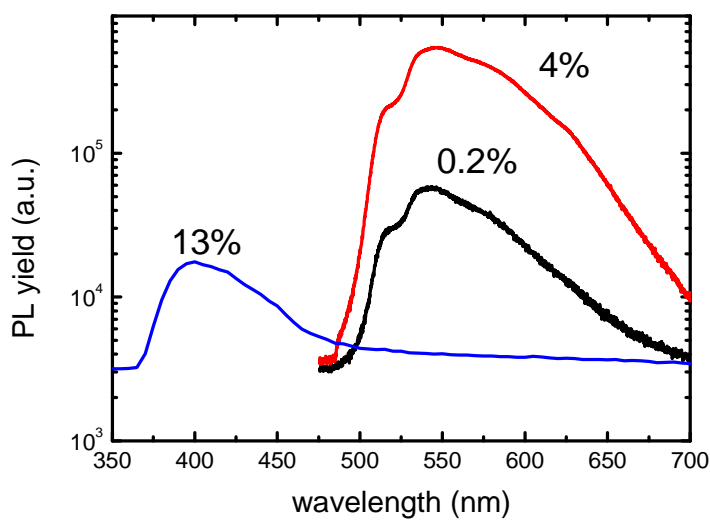

**Figure 4:** 5d -> 4f Ce transition probed by photoluminescence of the YAG nanoparticles doped with Ce at 0.2%, 4% and 13% from top to bottom, respectively. The emission of the 13% doped Ce:YAG nanoparticles is marginal and not representative of the samples which most parts do not emit light. The relative intensities are arbitrary and have been set only for the sake of clarity. Contrary to the 0.2% and 4% doped nanoparticles, which have been excited by a 450 nm wavelength, the 13% doped nanoparticles have been excited using a 220 nm wavelength.

Quantitative studies are complicated to achieve in photoluminescence, since the detected luminescence intensity depends on the absorption coefficient, which depends linearly on the cerium content and on the number of particles under the excitation beam. We have therefore developed a method combining the comparative CL measurements of the emission yield with the determination of the quantity of emitting matter as probed by RBS. It is clear that the measured luminescence efficiency does not represent exactly the same as under light excitation since promoting the cerium in the excited states is obtained through a complicated energy relaxation process of the interacting primary electrons. For a unit mass of YAG, the 4% doping leads to an increase in the luminescence yield of 31 (±3) times with respect to the 0.2% doping level. This already confirms that it is highly desirable to use high doping levels when using Ce:YAG nanoparticles in low quantity as individual nano-emitters. The increase in the luminescence yield is not proportional to the increase in the amount of Ce ions embedded in the particles (a given YAG nanoparticle doped at 4% with Ce contains 20 times more Ce ions that the same nanoparticle doped at 0.2%) since the emission of light is the result of a competition between energy transfers from the relaxed electron pairs, which remain in the particle after the electron cascade, toward the cerium in the trivalent states and toward the various quenching centres such as surface states.

The presence of such defects is indirectly observable in fluorescence decay-time experiments. Figure 5 presents the PL decays of the Ce:YAG nanoparticles doped at 0.2% and 4%. The decay of both samples can be fitted by a bi-exponential function (see Table 2 for the fitting parameters) indicating the existence of two distinct populations of Ce ions among the particles. In each case, a prominent contribution with a very fast decay time of about 3 ns is observed. This contribution, characterized by a decay time much smaller than that of $Ce^{3+}$ ions in a perfect bulk YAG crystal (~65 ns) [1], is clearly indicative of Ce ions in a strongly perturbed environment and in competition with very rapid quenching centres. Since this contribution is quite similar for the two samples, we can imagine that it originates from Ce ions in the same kind of environment. However, the two sets of samples differ by their second contribution. For the 0.2% doped nanoparticles, the second contribution is characterized by a short decay time of 9.7 ns whereas for the 4% doped nanoparticles, it exhibits a longer decay time of 27.7 ns. This value gets closer to that of the bulk crystal meaning that the corresponding Ce ions population is in a less perturbed environment.

Consequently, the main statement is that, in accordance with the result of the luminescence efficiency, the overdoping at 4% is optimal among the tested concentrations. It looks as if the doping Ce ions were initially and preferentially located in perturbed regions, where they are in competition with rapid quenching centres. When extra Ce ions are incorporated, they would occupy less perturbed sites. One hypothesis would be to consider that the first population corresponds to Ce ions at the surface of the particles. We have indeed demonstrated recently [24] that rare earth $Eu^{3+}$ ions embedded in a $Gd_2O_3$ nanoparticle exhibit significantly different emissions depending on their location within the particle. In particular, the emitting ions located within a 1nm thick shell at the surface have their emission strongly modified. On the contrary, the emitting ions located in the remaining core have an emission similar to the bulk one. Based on this statement, we propose that the two Ce ions populations observed in the present study may correspond to this core-shell distinction. If so, it would mean that in the present synthesis, the Ce ions tend to locate preferentially in the shell and only when their concentration is high enough do they occupy core sites.

**Table 2: values of the parameters from the fitting procedure of the PL decay curves of the Ce:YAG nanoparticles doped at 0.2% and 4%. $A_i$ is the amplitude of the $i^{th}$ contribution**

|  | $A_1$ (%) | $\tau_1$ (ns) | $A_2$ (%) | $\tau_2$ (ns) | $R^2$ |
|---|---|---|---|---|---|
| YAG:Ce 0.2% | 68 | 2.9 | 32 | 9.7 | 0.999 |
| YAG:Ce 4% | 80 | 3.6 | 20 | 27.7 | 0.998 |

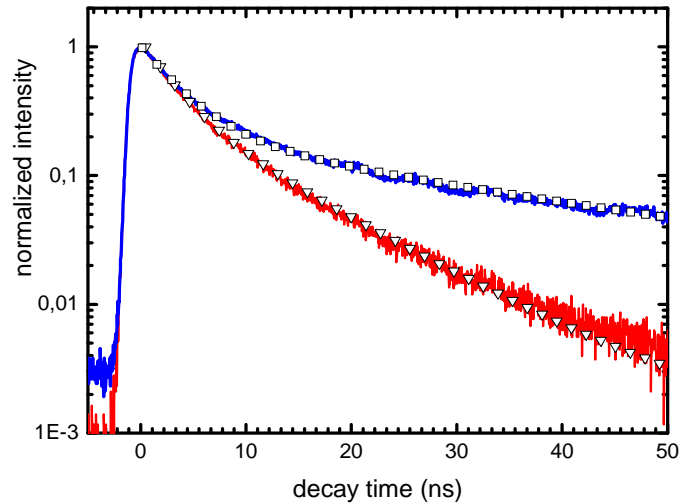

**Figure 5: PL decay of the Ce:YAG nanoparticles doped at 0.2% and 4%. The triangle and square symbols are fits to the decay curves of the 0.2% and 4% doped nanoparticles, respectively.**

The presence of quenching centres and defects being established we have sought to know if their presence was homogeneously distributed or in other words if the emission yields presented significant changes at the nanoscale. It is established that Ce tends to segregate at grain boundaries in bulk YAG ceramics [25].

To check this hypothesis, we have mapped the CL intensity of the 4% doped Ce:YAG nanoparticles film. Figure 6 shows the SEM image of a selected area together with the corresponding map of the CL intensity. Even though the CL intensity is rather noisy, its variation follows mainly that of the amount of matter in the studied assembly (along the dotted line). The intensity seems to be roughly proportional to the quantity of matter probed by the beam, going to show that the luminescent $Ce^{3+}$ ions as well as the quenching centres are not segregated on distances of the order of 50 nm or more. This is consistent with the fact that the quenching centres and Ce ions are likely distributed at the nanoparticles interfaces and inside the particles respectively, therefore on much shorter distances.

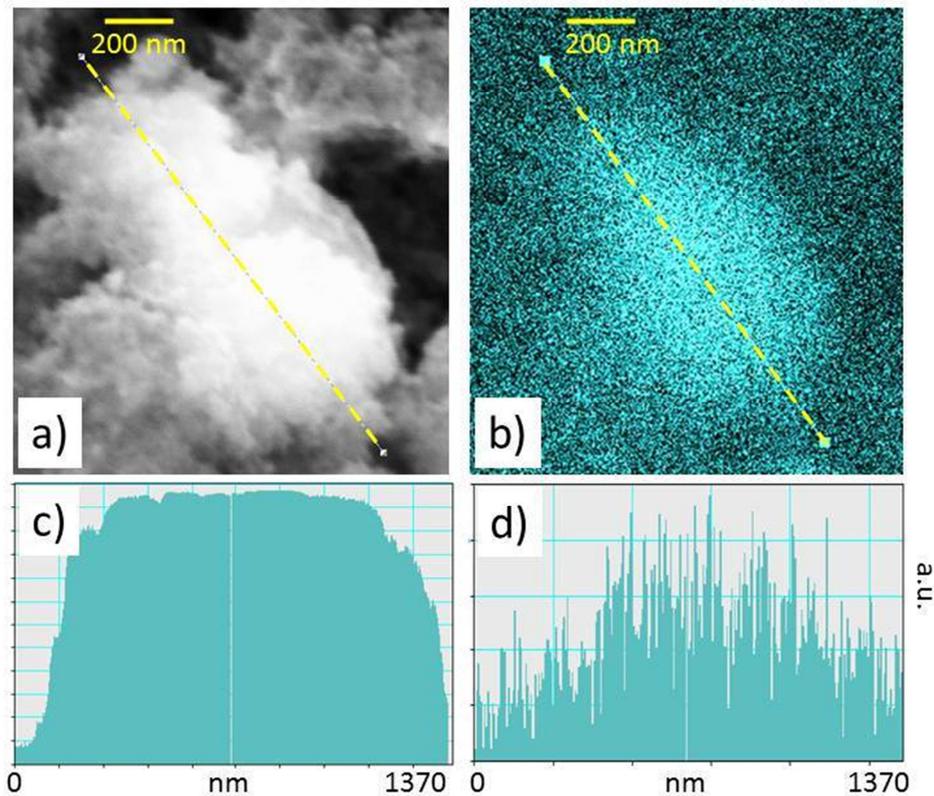

**Figure 6: a) SEM image of an assembly of 4% doped Ce:YAG nanoparticles. The foam-like structure of the film resulting from the assembly of nanoparticles is clearly visible; b) CL intensity map of the identical region. The scale bar is 200 nm. c) Variation of the quantity of matter (SEM signal intensity) along the dotted line visible in the SEM image a). d) Variation of the CL intensity along the same line. The intensity is mainly related to the quantity of emitting matter.**

4. Conclusion

We have synthesized by a physical method 10 nm Ce doped YAG nanoparticles with Ce concentrations of 0.2%, 4% and 13%. In the perspective to use these objects for the design of nano-emitters for photonic systems, we have avoided any subsequent post-deposition treatment and annealing that can be incompatible with the aimed applications. The structural analysis, based on TEM and XRD spectroscopy, reveals the amorphous nature of the particles assembled films. The composition of the particles obtained by XPS and RBS shows a slight defect in the stoichiometry, which may be responsible for the amorphous state of the particles. Despite this fact, the chemical environment (valence state) of the constituting elements is identical to the crystalline YAG one. The 13% doped particles contain both $Ce^{3+}$ and $Ce^{4+}$ ions and scarcely exhibit any emission, indicating that the amount of Ce ions is too large to be efficiently incorporated in YAG particles. On the other hand, the 0.2% and 4% doped particles have an emission spectrum identical to the bulk one. By a unique procedure combining CL and RBS we have measured the emission yield per unit mass under electron beam excitation. We first have confirmed that increasing the Ce concentration is highly beneficial for use as nano-emitters even with quenching; all the more so as the particles are amorphous. The presence of quenching centres is confirmed by time-resolved PL. A fast contribution in the decay curves is observed for both the 0.2% and 4% doped samples. Since this contribution does not depend on the Ce concentration, we regard it as "intrinsic" signature of our particles related to the crystalline disorder. A second component with a longer decay time of 9.7 ns and 27.7 ns is observed only for the 0.2% and 4% doped nanoparticles, respectively. This demonstrates that two distinct populations of Ce ions are present among the particles, one related to a strongly perturbed local environment, and another one corresponding to a local environment closer to the bulk one. Eventually, nano-CL does not reveal significant variations at the scale of 50 nm or larger of the emission yield and thus of the distribution of the emitting Ce ions as well as of the quenching centres.

**Acknowledgment**

The authors are indebted to Dr. D. Tainoff for TEM image acquisition and to Pr. R. Palmer (NPRL, university of Birmingham) for lending the nano-CL module. This work has been supported by the French National Agency (ANR) in the frame of its program in Nanosciences and Nanotechnologies (NAPHO project n°ANR-08-NANO-054).

[1] Liu G, Jacquier B 2005 *Spectroscopic Properties of Rare Earths in Optical Materials* (Springer Series in Material Science,)
[2] Yang X B, Li H J, Bi Q Y, Su L B, Xu J 2009 *J. Cryst. Growth* **311** 3692
[3] Zych E, Brecher C, Wojtowicz A J, Lingerat H 1997 *J. Lum.* **75** 193
[4] Yang H, D. Lee, Y. Kim, Mat. Chem. Phys. **114** 665 (2008)
[5] Schlotter P, Baur J, Hielscher C, Kunzer M, Obloh H, Schmidt R, Schneider J 1999 *Mat. Sci. Eng. B* **59** 390
[6] Haranath D, Chander H, Sharma P, Singh S 2006 *Appl. Phys. Lett.* **89** 173118
[7] Revaux A, Dantelle G, Decanini D, Haghiri-Gosnet A-M, Gacoin T, Boilot J-P 2011 *Opt. Mat.* **33** 1124
[8] Cuche A, Masenelli B, Ledoux G, Amans D, Dujardin D, Sonnefraud Y, Melinon P, Huant S 2009 *Nanotechnology* **20** 015603
[9] Purwanto A, Wang W-N, Ogi T, Lenggoro I W, Tanabe E, Okuyama K 2008 *J. All. Comp.* **463** 350
[10] Poitevin A, Chadeyron G, Briois V, Mahiou R 2011 *Mat. Chem. Phys.* **130** 500
[11] Su L T, Tok I Y, Boey F Y C, Zhang X H, Woodhead J L, Summers C J 2007 *J. Appl. Phys.* **102** 083541
[12] Kasuya R, Kawano A, Isobe T, Kuma H, Katano J 2007 *Appl. Phys. Lett.* **91** 111916
[13] Chevalier N, Nasse M J, Woehl J C, Reiss P, Bleuse J, Chandezon F, Huant S 2005 *Nanotechnology* **16** 613
[14] Cuche A, Mollet O, Drezet A, Huant S 2010 *Nano Lett.* **10** 4566
[15] Perez A. *et al.* 2010 *Int. J. Nanotechnol.* **7** 523
[16] Masenelli B, Nicolas D, Melinon P 2008 *Small* **4** 1233
[17] Nicolas D, Masenelli B, Mélinon P, Bernstein E, Dujardin C, Ledoux G, Esnouf C 2007 *J. Chem. Phys.* **125** 171104

[18] Masenelli B, Melinon P, Nicolas D, Bernstein E, Prevel B, Kapsa J, Boisron O, Perez A, Ledoux G, Mercier B, Dujardin C, Pellarin M, Broyer M 2005 *Eur. Phys. J. D* **34** 139
[19] Pawlak D A, Woźniak K, Frukacz Z, Barr T L, Fiorentino D, Seal S. 1999 *J. Phys. Chem. B* **103** 1454
[20] Lu C H, Jagannathan R 2002 *Appl. Phys. Lett.* **80** 3608
[21] Kasuya R, Isobe T, Kuma H 2006 *J. All. Comp.* **408-412** 820
[22] Teterin Y A, Teterin A Y, Lebedev A M, Utkin I O 1998 *J. Elec. Spect. Rel. Phenom.* **88-91** 275
[23] Datta P, Majewski P, Aldinger F 2009 *Mat. Charac.* **60** 138
[24] Masenelli B, Ledoux G, Amans D, Dujardin C, Mélinon P 2012 *Nanotechnology* **23** 305706
[25] Zhao W, Anghel S, Mancini C, Amans D, Boulon G, Epicier T, Shi Y, Feng X Q, Pan Y B, Chani V, Yoshikawa A 2011 *Opt. Mat.* **33** 684